\documentclass{article}
\usepackage{amssymb}
\usepackage{amsfonts}
\usepackage{sw20ssur}

\newtheorem{theorem}{Theorem}
\newtheorem{acknowledgement}[theorem]{Acknowledgement}

\newtheorem{conjecture}[theorem]{Conjecture}

\input{tcilatex}

\begin{document}

\title{Cosmology with Interacting Dark Energy}
\author{Manasse R. Mbonye$^{1,2}$ \\
%EndAName
\ \ \\
$^{1}$\textit{Michigan Center for Theoretical Physics}\\
\textit{Physics Department, University of Michigan}\\
\textit{2477 Randall Laboratory, Ann Arbor, MI 48109}\\
mbonye@umich.edu\\
\\
$^{2\thanks{%
\ \ New address}}$\textit{Department of Physics, Rochester Institute of
Technology}\\
\textit{84 Lomb Memorial Drive, Rochester, NY 14623-5604}\\
mrmsps@rit.edu\\
}
\maketitle

\begin{abstract}
The early cosmic inflation, when taken along with the recent observations
that the universe is currently dominated by a low density vacuum energy,
leads to at least two potential problems which modern cosmology must
address. First, there is the old \textit{cosmological constant problem},
with a new twist: the \textit{coincidence problem}. Secondly, cosmology
still lacks a model to predict the observed current cosmic acceleration and
to determine whether or not there is a future exit out of this state (as
previously in the inflationary case). This constitutes (what is called here)
a \textit{dynamical problem}. In this article a framework is proposed to
address these two problems, based on treating the cosmic background vacuum
(dark) energy as both dynamical and interacting. The universe behaves as a
vacuum-driven cosmic engine which, in search of equilibrium, always
back-reacts to vacuum-induced accelerations by increasing its inertia
(internal energy) through vacuum energy dissipation. The process couples
cosmic vacuum (dark) energy to matter to produce future-directed
increasingly comparable amplitudes in these fields by setting up
oscillations in the decaying vacuum energy density and corresponding
sympathetic ones in the matter fields. By putting bounds on the relative
magnitudes of these coupled oscillations the model offers a natural and
conceptually simple channel to discuss the \textit{coincidence problem},
while also suggesting a way to deal with the \textit{dynamical problem}. A
result with important observational implications is an equation of state $%
w\left( t\right) $ which specifically predicts a variable, quasi-periodic,
acceleration for the current universe. This result can be directly tested by
future observational techniques such as SNAP.
\end{abstract}

\section{Introduction}

In the last few years, evidence has mounted suggesting that the universe is
currently accelerating. Measurements of type Ia supernovae [1] indicate that
the evolution of the Hubble parameter departs from that expected for a
matter dominated universe, and behaves as if under the influence of a
negative pressure due to a smooth and dominant background (dark) energy.
Further evidence for a vacuum-dominated universe comes from a combination of
observations: large scale structure (LSS) [2] suggests a low matter density
universe while the cosmic microwave background (CMB) anisotropy data [3]
shows the density of the universe to be virtually critical, consistent with
the requirements of the early inflationary scenario [4]. Success of
inflation required that the universe be in a vacuum-dominated state with a
large associated potential energy. This feature, when taken together with
the observations of a currently low density vacuum energy dominating the
universe, leads to at least two potential problems that modern cosmology
must address. First, the discrepancy between the initial high potential
energy state and the current low background dark energy constitutes what has
been called [5] the Cosmological Constant Problem. A new twist to this
problem, and which has been called the \textit{coincidence problem}, relates
to the observation [1] that the current background vacuum energy density, $%
\rho_{v\text{ }}^{0}$ $\sim10^{-30}\ g\,cm^{-3}$ is not only not zero, but
is also of the same order of magnitude as the current density $\rho_{m}^{0}$
of the matter fields.

This state of affairs gives rise to yet another complication. The observed
current vacuum-dominated state of the universe is not \textit{a priori}
predicted by theoretical cosmology, and (as of now) its origin remains
mysterious.\ To this extent, it is not known whether or not there is a
future (graceful) exit from the resulting cosmic acceleration (as was the
case in the early universe). The situation reflects our current ignorance
with regard to the future dynamical evolution of the universe. It
constitutes a\textit{\ dynamical problem}.

Several approaches have been developed to address the \textquotedblleft why
is $\Lambda$ small now?\textquotedblright\ part of the Cosmological Constant
Problem. They include dynamical $\Lambda-$term models [6, 7], dynamical
equation of state models [8], and rolling scalar field models [9]. A common
feature in these models is that the vacuum energy takes on a dynamical
character, decaying from some initially large value to a small one. One
approach, usually favored by particle physicists, is to seek for mechanisms
to cancel or at least suppress $\Lambda$, as in Coleman's wormhole approach
[10] or by introducing compensating fields in the Lagrangian. In his notable
works, Brandenberger has argued (for a review see [11]) that gravitational
back-reaction may lead to a dynamical cancellation mechanism for a bare
cosmological constant by giving rise to an evolving anti-de Sitter-like
field.

Meanwhile, the observation, [1] of a small but non-vanishing background dark
energy has made the cosmological constant problem even worse, since it is
harder to advance a symmetry argument that justifies an \textit{almost}
non-vanishing cosmological constant. In 1998, Caldwell, Dave and Steinhardt
[12] proposed the existence of a background field with a dynamical equation
of state $w\left( t\right) $. They called it \textit{quintessence}. Since
then, tracker models based on [12] have evolved to discuss the possible
origin of the observed current cosmic acceleration. The issue of tying
together the disappearance of the early universe's cosmological constant
with what the source of the current dark energy might be, and why the
density of the latter is currently coincidentally close to that of the
matter fields, is a challenge for which cosmology still searches for a
physically justifiable solution. Some authors have suggested an anthropic
principle [13] to explain the observations.

In this article a potential framework for addressing both the Cosmological
Constant Problem and the \textit{dynamical problem} is proposed. The
framework is based on the premise that the background dark energy is both
dynamical and interacting. It is possible that there may be several physical
arguments which can be put forward to justify this notion, among which is
the following example. According to general relativity a vacuum dominated
universe (spacetime) has an asymptotically de Sitter geometry with an
asymptotically de Sitter horizon. Quantum considerations indicate (for a
review see e.g. [14a,b])\textbf{\ }that near the horizon such a spacetime is
associated with time asymmetric quantum states (analogous to those in the
Unruh vacuum) which lead to a radiation flux. This means that one can treat
the horizon (size $\chi^{-1})$ as a radiative surface with a horizon
temperature $T=\frac{\chi }{2\pi}$ and area $A_{\chi}=4\pi\chi^{-2}$ so that
the cosmological constant $\Lambda=3\chi^{2}$ which is the energy source $%
\frac{\Lambda}{8\pi G}$ in a de Sitter spacetime also becomes a source of
radiation flux, (i.e. particle creation). With this view, one is led to the
conclusion that the vacuum energy associated with $\Lambda$ must (1) decay
and (2) create matter in the process. It is in this respect that such cosmic
background vacuum (dark) energy is both dynamical and interacting. The issue
of a time dependant "cosmological constant" has always been a touchy one in
both General Relativity and Quantum Cosmology for seemingly different, but
actually, related reasons. In Quantum Cosmology, which is based on Quantum
Field Theory, there is no mechanism to change the (de Sitter) vacuum states;
while in General Relativity, the Einstein field equations demand that the
energy momentum tensor of the de Sitter spacetime can only be proportional
to the spacetime metric $g_{\mu\nu}$. On the other hand the foregoing
considerations suggest that just as the eternal black hole (and its
associated Hartle-Hawking vacuum) in classical General Relativity is an
idealization, analogously, a vacuum dominated universe can not stay in an
eternally fixed de Sitter state. As pointed out in [14b], the temperature
and entropy for a collapsing black hole and those of an asymptotically de
Sitter spacetime arise in identical manner due to identical mathematical
formalism. It would therefore be surprising if one spacetime (de Sitter) is
static while the other one is non-static. Consequently, in our treatment, $%
\Lambda$ must decay. Such a feature of an asymptotically de Sitter
spacetime, is among potential tools one can apply to our universe to
understand why "$\Lambda$ is small now but not zero" and also (as we find
soon) points to a potential resolution of both the \textit{coincidence
problem} and the \textit{dynamical problem.}

This paper discusses only the macroscopic effects (of the above underlying
microscopic physics of vacuum matter interactions), namely, the global
cosmic dynamics. The approach is reminiscent of what thermodynamics is to
statistical mechanics. In this respect the universe can be treated as a
vacuum-driven cosmic engine. The cosmic vacuum energy acts as the fuel,
doing work on the universe by accelerating it. As is the case with any
engine, it is impossible for the universe to convert all its fuel to
mechanical work without dissipating some through irreversible processes that
increase its internal energy. This is analogous to a statement of the first
law of thermodynamics and is suggestive of a first law of cosmic dynamics.
In turn, the irreversible dissipations lead to a cosmic form of the second
(or entropy creation) law\footnote{%
It is interesting that, in cosmology, while the universe is known the second
(entropy) law, the notion of a first cosmic law is not developed}, $%
s_{\;;\mu}^{\mu}\geq0$. Looking at the simplicity of these notions, it is
tempting to wonder what new feature the approach brings. In this treatment,
the irreversible dissipation process introduced above is, indeed, cardinal
to the dynamics of the universe, and to underscore this we make the
following \textit{Cosmic Equilibrium Conjecture} (CEC):

\begin{conjecture}
The universe will increase its inertia (matter creation) as a back-reaction
to any (vacuum energy) influences tending to move it away from (dynamical)
equilibrium.
\end{conjecture}

Facilitating such processes, on a macroscopic scale, is a (bulk viscous)
creation pressure $\pi_{c}$, which arises as a back-reaction to the
spacetime acceleration. The role of this back-pressure is to build up
inertia to oppose the change (acceleration) which creates it, in the first
place. It does this by creating matter. This behavior behavior is
reminiscent of equilibrium-seeking systems in nature. For example, in
electromagnetic induction, Lenz's law [15] predicts a back emf; while in
QCD, quark confinement is explained in terms of asymptotic freedom [16]. It
is interesting that in this latter process the system will actually create
matter, when faced with external influences (forces) tending to shift it
from equilibrium.

In the current treatment, the vacuum-dissipation/matter-creation process
couples the vacuum to matter through a parameter $K$, which we constrain. In
this scenario, (increasing cosmic inertia) physical modes will be created
until the cosmic acceleration is offset and the creation pressure $\pi _{c}$
vanishes. As an example, this is how (in this model) the universe is able to
get out of inflation and become matter dominated. With no further immediate
need for creation, the existing matter fields in a given comoving volume
begin to dilute normally, (which is relatively faster). Such redshifting of
the matter fields, eventually, leads to a further vacuum domination which in
turn commences a new cycle of cosmic acceleration and the consequent
creation. As a result, the process sets up oscillations in the decaying
vacuum and corresponding sympathetic ones in the matter fields. These
coupled oscillations constrain the fields to track each other naturally. As
the universe evolves in time, successive periods of acceleration are
dominated by ever decreasing (decaying) energy densities. The resulting
back-reactions will similarly be increasingly weaker, producing
infra-red-shifting dominated modes. Each such acceleration will be weaker
and last longer than its predecessors. This would explain why the current
cosmic acceleration has lasted much longer than the early cosmic inflation.

These ideas are used in this paper to establish bounds on the relative
magnitudes of the evolving fields. It is in this sense that the model offers
a natural and conceptually simple channel to discussing the \textit{%
coincidence problem}, while also suggesting a way to deal with the \textit{%
dynamical problem}. The approach also independently relates the current
vacuum-dominated state of the universe to the early conditions by suggesting
inflation as a natural initial condition to the current classical dynamics.
At the same time, it provides a rationale for a natural exit to, not only
the inflationary scenario but also, the current acceleration. Lastly, the 
\textit{Cosmic Equilibrium Conjecture} (CEC) suggests a physically motivated
ansatz to what is otherwise a (time-old) philosophical question, namely,
\textquotedblleft why does the universe need and create
matter?\textquotedblright . The model has significant observational
consequences. The constructed equation of state $w\left( a\right) $ for the
interacting dark energy suggests a quasi-periodic acceleration which can be
directly tested by future experiments such as SNAP [17]. In this article,
the emphasis is to study the interactions between dark energy and matter
fields and how, in particular, they lead to a resolution of the Coincidence
Problem. To this end, it is sufficient to consider the behavior of the
relevant energy-momemtum tensor. This leads to the evolution of the fields $%
\rho _{v}\left( a\right) $ and $\rho _{m}\left( a\right) $\ with the scale
factor, $a$ on a fixed gravitational background configuration. A derivation
of the time evolution of the scale factor $a\left( t\right) $, which
requires solving the resulting gravitational field equations for the matter
fields in the presence of interacting dark energy, will be made in a
forthcoming article.

The rest of the article is organized as follows. In Section 2 the working
equations for an interacting vacuum energy are set up. Section 3 discusses
the evolution of the vacuum energy density in the presence of matter
creating processes. Section 4 lays out the framework for the evolution of
the matter fields in such an environment. A potential resolution to both the 
\textit{coincidence problem} and the \textit{dynamical problem} is presented
and observational tests are mentioned. Section 5 concludes the article.
Throughout the article the terms \textit{cosmic} \textit{vacuum} \textit{%
energy} and \textit{dark\ energy }are\textit{\ }used interchangeably to
describe a background cosmic energy that is both dynamical and interacting%
\textit{.}

\section{Interacting cosmic vacuum energy: working equations}

\subsection{Features}

In modeling an interacting vacuum, we take as the generator of cosmic vacuum
energy $\frac{\Lambda }{8\pi G}$, a dynamical cosmological parameter, $%
\Lambda $ of the form 
\begin{equation}
\Lambda \left( t\right) =m_{pl}^{4}\left( \frac{a_{pl}}{a(t)}\right)
^{\sigma \left( t\right) }e^{-\tau H}=\Lambda _{pl}\left( \frac{a_{pl}}{a(t)}%
\right) ^{\sigma \left( t\right) }e^{-\tau H},  \tag{2.1}
\end{equation}%
where $m_{pl}$ is the Planck mass, $a_{pl}$ (the fluctuation scale) is the
size scale of a causally connected region of space at the Planck time $t_{pl%
\text{, }}$and $\tau $ is of order of the Planck time. Further, $H$ is the
Hubble parameter and $a\left( t\right) $ is the cosmic scale factor. A
distinguishing feature of the model, is that the power index $\sigma $ is
not a fixed constant in time (as in most previous cases reviewed in [6]) but
rather depends on time in a manner to be determined in Section 3. This
feature, which is a consequence of the \textit{Cosmic Equilibrium Conjecture}
allows the background dark energy to couple to matter.

As is apparent from Eq. 2.1, in this scenario, the $\Lambda $-field
initially $\left( H^{-1}\rightarrow 0\right) $ appears to tunnel into
existence. Thereafter, for $t\lesssim \tau $ the dynamics of the early
universe is driven by the growing term $e^{-\tau H}$ and should be dominated
by a quantum character. A rigorous discussion of the physics of the period $%
t\lesssim \tau $ requires a quantum theory of gravity and is, therefore,
still beyond the scope of the present treatment. It is, however, worth
pointing out that for $t\lesssim \tau $ the field in Eq. 2.1 is growing in
density and so one expects its equation of state during this early period to
take the form $w\left( t\right) <-1$. As the field evolves towards its
stationary point, $\frac{d\Lambda \left( t\right) }{dt}=0$, the equation of
state $w\left( t\right) $ approaches $-1$ from below, to temporarily mimic a
cosmological constant. In the immediate neighborhood of $\frac{d\Lambda
\left( t\right) }{dt}=0$, or $\left[ \ln \left( \frac{l_{pl}}{a(t)}\right) 
\frac{d\sigma }{dt}-\sigma \left( t\right) H-\tau \frac{dH}{dt}\right] =0$, $%
\Lambda $\ is virtually constant with maximum potential energy (possibly in
the range $\langle \rho _{vac}\rangle \ =\frac{\Lambda }{8\pi G}\sim
\;10^{94}\,g\,cm^{-3}$). These conditions give rise to a cosmic inflation,
in the early universe$.$ As the Hubble time $H^{-1}$ grows, the quantity $%
e^{-\tau H}$ quickly approaches saturation $e^{-\tau H}\rightarrow 1$.
Subsequently, the dynamics of the universe becomes increasingly classical,
being driven by the $a(t)^{-\sigma \left( t\right) }$ part of $\Lambda $. It
is this latter phase that this article discusses.

Assuming successful inflation in the early era, then by the time the
universe is about $1$ second old, the scale factor is usually considered to
have grown by a factor of about $10^{28}$. In our case, this means that $%
\Lambda $ will have decayed by $10^{-28<\sigma \left( t\right) >}$, in the
process (leaving behind a dense field of relativistic particles and
radiation). It turns out (as is shown in Section 3), that the time average
value of the power index function, in the model, is $<\sigma \left( t\right)
>=2$. As a result, the cosmic vacuum energy density does not interfere with
the usual early cosmological processes like big bang nucleosynthesis (BBN)
[18], but instead allows such processes to proceed as predicted by the
standard big bang model. Further, since the inflation era to date, $a(t)$ is
known to have evolved by about $10^{60}$. The overall result is that $\left[
\Lambda _{now}\sim a_{0}^{-2}\right] \approx 10^{-120}\Lambda _{pl}$. This
provides a heuristic explanation to the \textquotedblleft why is $\Lambda $
small now\textquotedblright\ part of the cosmological problem, which is
consistent with observations, pending the proof (later in the discussion)
that $<\sigma \left( t\right) >=2$.

In the remaining part of this article, tools are developed to address both
parts of the Cosmological Constant problem and the \textit{dynamical problem}%
. Throughout the proceeding discussion, we only deal with the late-time
evolution of the universe. Here, $H^{-1}>>\tau$ so that $e^{-\tau H}$ can,
justifiably, be set to unity with the result that the effective late-time $%
\Lambda$ is controlled by $a\left( t\right) ^{-\sigma\left( t\right) }$ and
the universe evolves quasi-classically$^{2}$\footnotetext[2]{%
A matter creating universe can not be entirely classical.}. The associated
vacuum energy density can, thus, be written as 
\begin{equation}
\rho_{v}\left( t\right) =\frac{\Lambda\left( t\right) }{8\pi G}=\left( \frac{%
a\left( t_{0}\right) }{a\left( t\right) }\right) ^{\sigma\left( t\right)
}\rho_{v}^{\left( 0\right) },  \tag{2.2}
\end{equation}
where, for convenience (in the second equality) the dynamical vacuum energy
density is normalized about its observed current value $\rho_{v\text{ }}^{0} 
$ $\sim10^{-30}\ g\,cm^{-3}$. In our notation the sup/sub-script $0$ on a
quantity denotes its current value.

\subsection{Energy equations and particle creation}

Consider the dynamical evolution of a self-gravitating cosmic medium
consisting of a two-component perfect fluid. The total energy momentum
tensor $T_{\mu\nu}$ for all the fields is given by 
\begin{equation}
T_{\mu\nu}=T_{\mu\nu}^{\left( m\right) }+T_{\mu\nu}^{\left( vac\right) }= 
\left[ \rho+p\right] v_{\mu}v_{\nu}+pg_{\mu\nu},  \tag{2.3}
\end{equation}
where $T_{\mu\nu}^{\left( m\right) }$ and $T_{\mu\nu}^{\left( vac\right) }$
are, respectively, the matter and the vacuum contributions to $\ T_{\mu\nu }$%
, while $\rho=\rho_{m}+\rho_{v}$,\ $\ p=p_{m}+p_{v}$ and $v_{\mu}$ is the
4-velocity. Conservation of the total energy, $v_{\mu}T_{\;\;;\nu}^{\mu%
\nu}=0 $, leads to standard continuity equation 
\begin{equation}
\left[ \dot{\rho}_{m}+\left( \rho_{m}+p_{m}\right) \theta\right] +\left[ 
\dot{\rho}_{v}+\left( \rho_{v}+p_{v}\right) \theta\right] =0,  \tag{2.4}
\end{equation}
where $\theta=v_{\;;\alpha}^{\alpha}$ is the fluid expansion parameter and $%
\dot{\rho}$ is the derivative taken along the fluid worldline,$\ \dot{\rho }%
=v^{\alpha}\nabla_{\alpha}\rho$. In this treatment, the total energy in the
cosmic fluid is conserved. However, because of the assumed interacting
nature of dark energy the individual components are, in general, not
conserved. In particular, because of the CEC (and under conditions to be
discussed) the vacuum will act as a source of dissipative processes, while
the matter component acts as a sink of such processes. This means that one
can write 
\begin{equation}
v_{\mu}T_{\left( vac\right) \;\;;\nu}^{\mu\nu}=-v_{\mu}T_{\left( m\right)
\;\;;\nu}^{\mu\nu}=\Psi  \tag{2.5}
\end{equation}
where $\Psi>0$ is the particle source strength. Note that Eq. 2.5 is still
consistent with Eq. 2.4.

In principle, a non-equilibrium system can involve dissipative processes,
ranging from scalar fluxes like bulk viscous pressure and particle creation
pressure to tensorial shear viscosity stresses and energy transport [19].
For the case under consideration, the latter are globally suppressed in an
isotropic and homogeneous universe as assumed here. The contribution to the
entropy source, by the remaining scalar processes, is given by [19] 
\begin{equation}
S_{;\alpha }^{\alpha }=-\frac{\Pi \theta }{T}-\frac{\pi _{c}\theta }{T}-%
\frac{\mu \Psi }{T},  \tag{2.6}
\end{equation}%
where $\pi _{c}$ is the creation pressure, $\Pi $ is the bulk viscous
pressure, $\mu $ is the chemical potential and $T$ is the temperature. In
general, it can be shown [19, 22] that in the absence of the particle source
strength $\Psi $, the creation pressure and the bulk viscosity become the
same process and there is no particle creation.

In this article, we only pay attention to particle creating processes
described by the creation pressure, $\pi _{c}$. Then, use of Eq. 2.5 shows
Eq. 2.4 to consist of two dissipative equations 
\begin{equation}
\left[ \dot{\rho}_{v}+\left( \rho _{v}+p_{v}\right) \theta \right] =\pi
_{c}\theta  \tag{2.7}
\end{equation}%
and 
\begin{equation}
\left[ \dot{\rho}_{m}+\left( \rho _{m}+p_{m}\right) \theta \right] =-\pi
_{c}\theta .  \tag{2.8}
\end{equation}

We suppose that the interacting dark energy satisfies an effective equation
of state of the form 
\begin{equation}
p_{\Lambda }=w\rho _{\Lambda },  \tag{2.9}
\end{equation}%
where $-1\leq w\leq 0$. Note the `unusual' upper limit $w\leq 0$ (instead of 
$w\leq -\frac{1}{3}$) for the interacting dark energy, signifying a matter
dominated state, in which the effective pressure of the interacting dark
energy hardly contributes to the dynamics of the universe. The quantity $w$
will have an implicit time dependence (through the fields) when particle
creating processes are in force. It will be shown (see next sub-section)
that $w\longrightarrow -1$ in the limit\ $\rho _{m}\longrightarrow 0$. Thus
in this limit, one recovers (in an asymptotically locally inertial frame),
the standard Lorentz invariant vacuum. Further, we assume [20] that the
newly created particles are virtually in thermal equilibrium with the
existing matter fields as soon as they are created. This is reasonable in
light of the approach we have adopted above which suppresses non-matter
creating bulk viscosity effects. Thus, the only source of entropy is matter
creation. As a result, the matter fields satisfy the usual $\gamma $-law
equation of state 
\begin{equation}
p_{m}=\left( \gamma -1\right) \rho _{m},  \tag{2.10}
\end{equation}%
where $\gamma =\left\{ 1,\frac{4}{3}\right\} $.

The density fields $\rho _{v}$ and $\rho _{m}$ can be determined using any
two of the three equations, namely, the energy balance (continuity) equation
(Eq. 2.4) and the source-sink equations (Eqs. 2.7 and 2.8). To proceed,
however, one requires the functional forms of both the power index function $%
\sigma \left( a(t)\right) $ and the\ effective equation of state $w\left(
a\left( t\right) \right) $ for the interacting vacuum. This problem is
addressed in the next section.

\section{Evolution of vacuum energy density}

In this section we develop the functional form of the interacting vacuum
energy density $\rho _{v}$ and the pressure $p_{v}$ by deriving $\sigma
\left( a\right) $ and $w\left( a\right) $, and study how the evolution of
these fields suggests a natural resolution of the two problems set out in
Section 1.

\subsection{Creation pressure}

In a vacuum dominated universe, the total gravitating energy is $\rho +3p<0$
(which, as is known, violates the strong energy condition SEC). The excess
negative pressure accelerates the universe. In turn, according to the 
\textit{Cosmic Equilibrium Conjecture}, the universe back-reacts to this
non-equilibrium scenario by building inertia (matter modes) through the
creation pressure $\pi _{c}$.\ We assume the local equilibrium hypothesis
[21] that non-equilibrium quantities in the model depend locally on similar
variables as the equilibrium ones. It follows, then, that the particle
creating pressure $\pi _{c}$, which must depend on the available excess
negative pressure, will be proportional to the total gravitating energy, $%
\rho +3p$, of the universe. Consequently, we can write 
\begin{equation}
\pi _{c}=K\left[ \left( 3\gamma -2\right) \rho _{m}-2\rho _{v}\right] , 
\tag{3.1}
\end{equation}%
where the dimensionless proportionality parameter $K$ is to be constrained.
This parameter $K$ (here and henceforth referred to as the vacuum
dissipation parameter) couples the cosmic vacuum energy to matter through
vacuum dissipation/matter creation processes (Eqs. 2.7 and 2.8). In the
(idealized) limit $\rho _{m}\rightarrow 0$, $K$ would probe the efficiency $%
\epsilon =1-K$ of the universe as a cosmic engine. Our immediate goal is to
relate the creation pressure $\pi _{c}$ to the dynamical evolution of the
density fields $\rho _{v}$ and $\rho _{m}$. To proceed, we start by
constructing an effective equation of state for the interacting background
dark energy.

\subsection{Cosmic Expansion and Field dilution}

As the universe expands the densities of the background matter fields are
known to dilute as $a(t)^{-3\gamma}=a(t)^{-3(1+\breve{w})}$, where $\breve
{%
w}=\left\{ 0,\frac{1}{3}\right\} $. On the other hand, a dynamical
interacting vacuum energy (of the functional form in Eq. 2.2) will suffer an
energy deficit because it is a source of matter fields $v_{\mu}T_{\left(
vac\right) \;\;;\nu}^{\mu\nu}=\Psi\neq0$. The vacuum energy in a given
comoving volume will appear to redshift with a dilution law $a(t)^{-3(1+w)}$%
. To preserve the functional behavior of the vacuum energy density, in Eq.
2.2, one must have $3\left( 1+w\right) =\sigma\left( a\left( t\right)
\right) $, where the quantity $\sigma\left( a\left( t\right) \right) $ is to
be determined. This gives $w=-\left( 1-\frac{\sigma}{3}\right) $, which
leads to an effective equation of state%
\begin{equation}
p_{v}=-\left( 1-\frac{\sigma}{3}\right) \rho_{v}.  \tag{3.2}
\end{equation}

It is convenient (for now) to write the time evolution of the vacuum energy
as an evolution in the scale factor $a\left( t\right) $. In the FRW models,
the source of the four-velocity is the Hubble parameter, then $v_{\;;\alpha
}^{\alpha }=$ $\theta =3\frac{\dot{a}}{a}=3H$. This, along with use of Eq.
3.2, transforms Eq. 2.7\ to 
\begin{equation}
a\acute{\rho}_{v}+\sigma \rho _{v}=3\pi _{c},  \tag{3.3}
\end{equation}%
where $\acute{\rho}_{v}=\frac{d\rho _{v}}{da}$ and as before $\rho _{v}=\rho
_{v}^{0}\left( \frac{a_{0}}{a}\right) ^{\sigma \left( a\right) }$. Using
Eqs. 3.1, 3.2 and 3.3 the effective equation of state of an interacting dark
energy can be written as%
\begin{equation}
w=-1+\left\{ \frac{3K\left[ \left( 3\gamma -2\right) \rho _{m}-2\rho _{v}%
\right] -a\acute{\rho}_{v}}{3\rho _{v}}\right\} .  \tag{3.4}
\end{equation}

\subsection{\protect\smallskip Interacting dark energy: a solution}

One expects that the asymptotic forms of the relation in Eq. 3.4 should
recover the more familiar equations of state of ordinary physical fields. In
particular, when the vacuum energy dominates the universe $\left( \rho
_{v}>>\rho_{m}\right) $, one expects $w\rightarrow-1$. From Eq. 3.4, this
limit requires that $-2K-\frac{1}{3}\left( a\frac{\acute{\rho}_{v}}{\rho_{v}}%
\right) =0$, which integrates to 
\begin{equation}
\rho_{v}\mid_{\rho_{m}\rightarrow0}=\rho_{v}^{0}\left( \frac{a_{0}}{a}%
\right) ^{6K},\;\left( K\geq0\right) .  \tag{3.5}
\end{equation}

\smallskip On the other hand, the system evolves so that eventually there is
no interaction between the vacuum and matter fields once the creation
pressure $\pi_{c}$ vanishes. This leads to the limit where effectively $%
w\longrightarrow0$. \ Applying this limit on Eq. 3.4 gives $w=-1-\frac{1}{3}%
\left( a\frac{\acute{\rho}_{v}}{\rho_{v}}\right) =0$, so that 
\begin{equation}
\rho_{v}\mid_{\pi_{c}\rightarrow0}=\rho_{v}^{0}\left( \frac{a_{0}}{a}\right)
^{3}.  \tag{3.6}
\end{equation}
The above limits (Eqs. 19 and 20) establish bounds on the power index
function $\sigma\left( t\right) $ as$\ $ 
\begin{equation}
6K<\sigma\leq3  \tag{3.7}
\end{equation}
At $\sigma=3$, the vacuum (dark) energy has a maximum dilution rate and
decouples from matter. The evolution of the dark energy to this state also
implies a relative increase (and eventual domination) of the matter fields
over the vacuum energy. Here, matter creation is suppressed consistent with
the requirements of the \textit{Cosmic Equilibrium\ Conjecture}. With no
more creation, the relative matter dilution grows towards its
\textquotedblleft normal\textquotedblright\ rate of $\sim a^{-3\gamma}$. In
turn, this increases the density of the dark energy relative to that of the
matter fields, eventually setting the former into dominance. The universe
then begins to accelerate and, in the process, builds the creation pressure $%
\pi_{c}$, as a back-reaction to the acceleration. According to Eqs. 3.5 and
3.7 this creation rate will grow towards a maximum, as $\sigma$ approaches
its minimum, $\sigma_{\min}=6K$. As the matter creating vacuum dumps in more
and more matter, the matter fields are evolving with a decreasing dilution
rate. The relative increase in the density of matter fields, makes it less
and less favorable for the vacuum to create more matter. As a result, the
creation rate is decreasing towards a minimum. Eventually, the system ends
up back where it started (with no more creation) as $\rho_{m}\sim
a^{3\gamma} $\ and $\pi _{c}\sim0$\ and a new cycle begins.

It follows, then that, in general, the power index function $\sigma \left(
t\right) $ will be oscillatory within the bounds $6K<\sigma \left( t\right)
\leq 3$ as established in Eq. 3.7. Oscillations in $\sigma \left( t\right) $
naturally imply oscillations in the decay rate of the density function, $%
\rho _{v}\sim a^{-\sigma \left( t\right) }$. Here, it should be pointed out
that while the power index function (we seek) is oscillatory, the density
function $\rho _{v}\sim a^{-\sigma \left( t\right) }$ must be single-valued
in $a\left( t\right) $ in order to be consistent with the requirement that
(globally) matter/entropy creation from the vacuum is an irreversible
process. The vacuum energy decays into matter but not vise versa. This
concept is explored further in Section 3.4.1. Thus such oscillations will be
imprinted on a decaying energy background. As is shown below, this feature
is inherent in the model.

The oscillations in the decay rate of $\rho_{v}$ signify matter creation
from the vacuum. This implies that such periodic matter creation will
necessarily induce sympathetic oscillations in the matter density fields $%
\rho_{m}$. It is in this sense that the two fields, in time, track each
other. To support these assertions we start by studying how the creation
pressure will drive the oscillations.

Recall $\rho _{v}=\left( \frac{a_{0}}{a}\right) ^{\sigma \left( t\right)
}\rho _{v}^{0}$. On taking derivatives with respect to $a\left( t\right) $,
we have that $\acute{\rho}_{v}\left( a\right) =\left[ \acute{\sigma}\ln
\left( \frac{a_{0}}{a}\right) -\frac{\sigma }{a}\right] \rho _{v}$.
Substituting for $\acute{\rho}_{v}$ in Eq. 3.3 gives 
\begin{equation}
\acute{\sigma}a\ln \left( \frac{a_{0}}{a}\right) =\frac{3\pi _{c}}{\rho _{v}}%
.  \tag{3.8}
\end{equation}%
Now, consider $\sigma \left( \psi \right) $, where $\psi $ is some function
of $a\left( t\right) $. Then $\acute{\sigma}\frac{da}{d\psi }=\frac{d\sigma
\left( \psi \right) }{d\psi }$. Comparing this with Eq. 3.8 one sees that,
on setting $\frac{d\psi }{da}=\left[ a\ln \left( \frac{a_{0}}{a}\right) %
\right] ^{-1}$, then $\ $%
\begin{equation}
d\sigma =\left( \frac{3\pi _{c}}{\rho _{v}}\right) d\psi .  \tag{3.9}
\end{equation}%
Thus, $\frac{3\pi _{c}}{\rho _{v}}d\psi $\ is a perfect differential of the
power index function $\sigma $.

One expects solutions to Eq. 3.9 with certain specific characteristics.
First, from the discussion above, such solutions should be oscillatory in $%
\psi$\ and also bounded by Eq. 3.7. Secondly, all the $\psi\left( a\left(
t\right) \right) $ dependence in the solution $\sigma$ should be purely
sinusoidal, so as to avoid unphysical solutions of the form $\rho\sim
a^{f\left( a\right) }$, where $f\left( a\right) $ is non-oscillatory.
Finally, as is pointed out in the \textit{Cosmic Equilibrium Conjecture},
matter creation (in this model) is an opposite reaction to the
vacuum-induced \textit{positive} acceleration of the universe. Thus, the
sinusoidal part of $\frac{d\sigma}{d\psi}$ should be negative definite, i.e.
of the form $\sim\left( -\sin^{2}\psi\right) $, at the least.

The simplest solution that satisfies these conditions has the form $\sigma
=\sin 2\psi +A$, where $A$\ is a positive constant. Only then are the
requirements on $\frac{d\sigma }{d\psi }$ satisfied, since now $\left\{ 
\frac{d\sigma }{d\psi }=\frac{3\pi _{c}}{\rho _{v}}=3K\left[ \frac{\left(
3\gamma -2\right) \rho _{m}-2\rho _{v}}{\rho _{v}}\right] \right\} =-4\sin
^{2}\psi +2$. On rearranging out the terms in the last equality of this
equation one finds that 
\begin{equation}
\sin ^{2}\psi =\left[ \frac{2\left( 1+3K\right) \rho _{v}-3K\left( 3\gamma
-2\right) \rho _{m}}{4\rho _{v}}\right] .  \tag{3.10}
\end{equation}%
This function is minimum $\left( \sin ^{2}\psi =0\right) $ at 
\begin{equation}
_{\min }\rho _{v}=\left( \frac{3K}{6K+2}\right) \left( 3\gamma -2\right)
\rho _{m},  \tag{3.11}
\end{equation}%
and maximum $\left( \sin ^{2}\psi =1\right) $ at 
\begin{equation}
_{\max }\rho _{v}=\left( \frac{3K}{6K-2}\right) \left( 3\gamma -2\right)
\rho _{m}.  \tag{3.12}
\end{equation}%
We soon return to these limits in the next section. To formally complete the
solution $\left( \sigma =\sin 2\psi +A\right) $, one notes on applying the
limits from Eq. 3.7 that when $\sin 2\psi =1$, then $1+A=3$. With this, we
find the solution for the power index function $\sigma $ to be 
\begin{equation}
\sigma \left( \psi \right) =2+\sin 2\psi ,  \tag{3.13}
\end{equation}%
where as shown earlier, $\psi $ is given by $\frac{d\psi }{da}=\left[ a\ln
\left( \frac{a_{0}}{a}\right) \right] ^{-1}$. Using Eq. 3.13 in Eq. 2.2 we
finally find that the interacting dark energy density, in this model,
evolves with the scale factor as 
\begin{equation}
\rho _{v}\left( a\right) =\left[ \frac{a_{0}}{a}\right] ^{\left( 2+\sin
2\psi \left( a\right) \right) }\rho _{v}^{0}.  \tag{3.14a}
\end{equation}%
or in terms of the redshift parameter $z=\frac{a_{0}}{a}-1$ we have 
\begin{equation}
\rho _{v}\left( z\right) =\left[ z+1\right] ^{\left( 2+\sin 2\psi \left(
z\right) \right) }\rho _{v}^{0},  \tag{3.14b}
\end{equation}%
where now $\frac{d\psi }{dz}=\left[ \left( z+1\right) \ln \left( z+1\right) %
\right] ^{-1}$. Further, Eqs. 3.2 and 3.13 give the working equation of
state for the interacting dark energy in this model 
\begin{equation}
p_{v}=-\frac{1}{3}\left( 1-\sin 2\psi \right) \rho _{v}.  \tag{3.15}
\end{equation}%
It is this equation of state that, in future, can be compared with
observations like SNAP [17].

\subsection{Some features of the $\protect\rho_{v}$ solution}

Eqs. 3.13 and 3.15 give the formal solution for the interacting dark energy
in this model. According to this solution, during the evolution of the
universe, $\rho_{v}$ has the highest dilution rate at points characterized
by $\sin 2\psi=1$, where $\psi_{n}=\left( \frac{1}{4}+n\right) \pi,\ \left\{
n=0,1,2,...\right\} $. It is here that the vacuum energy density becomes
least interacting, since effectively it decouples from the matter fields as $%
w\longrightarrow0$. On the other hand, the field has its least dilution rate
(and is most interacting) at epochs characterized by $\psi_{n}=\left( \frac{3%
}{4}+n\right) \pi,\;\left\{ n=0,1,2,...\right\} $.

\subsubsection{Mean decay path and why $\Lambda $ is small now}

For $\psi _{n}=\frac{n}{2}\pi ,\;\left\{ n=0,1,...\right\} ,\ \sin 2\psi ~=0$%
. Since $\psi =\psi \left( a\left( t\right) \right) $, and noting that $%
<\sin 2\psi >~=0$, we see from Eq. 3.13 that $<\sigma >\,=2$. The
implication here is that the interacting vacuum energy density, in this
model, decays along a mean evolutionary path of $a^{-2}$, with oscillations
of $\sin 2\psi $ about this decay path. Thus, in an expanding universe,
according to this result, the interacting vacuum energy density is
inherently a decaying system. This is, indeed, consistent with our premise
(in Section 1) that $\Lambda $ must decay. In Section 2.1, a heuristic
argument was given to explain why $\Lambda $ is small now. It was seen that
setting $\sigma =2$ implies $\left[ \Lambda _{now}\sim a_{0}^{-2}\right]
\approx 10^{-120}\Lambda _{pl}$. In our treatment, we have recovered this
value as the average of the oscillations in the power index, $<\sigma >\,=2$%
. Thus our treatment solves the "why is $\Lambda $ small now" in a way that
is consistent with current observations [1].

In the past, several phenomenological models have been developed (see [6]
and citations) in which the vacuum energy density $\rho _{v}$ evolves with
the scale factor as $a^{-m},$ where the index $m$ has a fixed value. In such
models $\rho _{v}$ is not explicitly interacting, as is indicated by the
constant nature of the index $m$. One such set of models that has gained
considerable popularity (see citations in [6]) evolves $\rho _{v}$ as an
inverse square power law ($m=2$), in the scale factor. For the reasons given
above, the ($m=2$) models [6] have in the past offered a satisfactory
resolution of why $\Lambda $ is small now. It is worthy noting that the
approach presented in the present work recovers this inverse square law as
the mean decay path for an interacting vacuum energy density, $<\rho _{v}>$.
Moreover, for $a>>a_{0}$, as the universe ages $\sigma \longrightarrow 2$ so
that (see also Section 4.2) $\sigma $ can be replaced by its average value, $%
<\sigma >\,=2$. In this way, our model can be related to the inverse square, 
$\Lambda \sim a^{-2}$, models.

The positivity feature of $\sigma$ (see Eq. 3.13), $\sigma\,\geq1>0$,
ensures that even in the face of the oscillations in the power index
function, the energy function $\rho_{v}$ is still always a decreasing
single-valued function of the scale factor $a\left( t\right) $. To
appreciate the physical significance of this feature, one need only consider
its contradiction, where $\rho_{v}$ evolves as a multi-valued function of $%
a\left( t\right) $. Such a scenario would imply a two-way process where the
vacuum (dark) energy and matter alternately turn into each other. Globally,
this would be unphysical (at least in our part of the universe) since it
would result in periodic global violations of the entropy (evolution) law
and suggest a rotating thermodynamic arrow of time. In this model, the
creation processes tending to increase cosmic inertia are one-way and lead
to entropy growth. Thus the treatment ensures that the evolutionary law of $%
\rho_{v}$ is consistent with a causal universe.

\subsubsection{The parameter space of $K$}

The preceding discussion depicts a universe undergoing periods of
vacuum-driven acceleration, punctuated by periods of matter domination. One
can gain some deeper insight in the cosmic dynamics during each of such
acceleration periods by paying attention to the associated parameter $K$.
Suppose, for example, at some point in time (not unique) during a given
period of cosmic acceleration one can measure the contemporary values of $%
\rho_{v}$ and $\rho_{m}$ or just the ratio $\beta=\frac{\rho_{v}}{\rho_{m}}$%
. Then the maximum value $\beta_{\max}$, coinciding with $_{\max}\rho_{v}$
during such a (local) period can not be lower, i.e. $\beta_{\max}\nless\beta$%
. This argument can be used to put reasonable bounds on $K$ during each such
period since from Eq. 3.13 we must have 
\begin{equation}
\beta\leq=\beta_{\max}=\left( \frac{3K}{6K-2}\right) \left( 3\gamma
-2\right) .  \tag{3.16}
\end{equation}
Thus any measured value of $\beta$ gives one information on the upper-bound
on $K$ for that period (which obviously would be better tightened by
knowledge of $\beta_{\max}$). For example, as observations indicate, the
current value is $\beta\approx2$. This suggests that for the current (cold $%
\gamma=1$) universe, $K\ngtr\frac{4}{9}$. Moreover\ from Eq. 3.10 one gets a
lower bound on $K$ for the interactions to result in particle production. In
the (idealized) limit $\rho_{m}\rightarrow0$, one finds that $%
K\longrightarrow \frac{1}{3}$.

The two bounds above thus constrain the parameter space of the dissipation
parameter $K$ in the current universe to 
\begin{equation}
\frac{1}{3}<K\lesssim\frac{4}{9}.  \tag{3.17}
\end{equation}

The one-way character of the dissipation process brings out yet another
important feature. Clearly as the universe gets older, the upper bound on $K$
grows as $\beta_{\max}\longrightarrow1$ revealing the source of the current 
\textit{coincidence problem}. In the next section we put bounds on their
relative amplitudes as the fields evolve. The above lower bound $\frac{1}{3}%
<K$ suggests that, in this model, the parameter space $0\leq K\leq\frac {1}{3%
}$ is forbidden with regard to matter creation. One can, however, imagine a
time in the very early history of the universe when the vacuum energy
therein was very large and in a purely potential $\left( PE\right) $ form.
Such conditions would imply that, at the time, $K=0$. Then Eq. 3.7 shows
that $\rho_{v}\sim a^{-\left( 6K\right) =0}$ and the vacuum energy density
is constant in time (albeit temporarily).\ Only then could the universe as
an engine seem to operate at $100\%$ efficiency. Clearly this is the energy
in a cosmological constant and it would inflate the universe. In turn,
because inflation would tend to strongly and suddenly move the universe away
from equilibrium conditions, the\textit{\ Cosmic Equilibrium Conjecture}
implies that the associated backreation also be strong and sudden. Thus,
virtually all the physical (matter) modes are created in the early universe.
The space $0\leq K\leq\frac{1}{3}$ during which matter creation is forbidden
seems to provide a small window for inflation before the backreaction sets
in as $K\longrightarrow0$ from below. Accordingly, there are two
consequences to this. First, it is in immediate reaction to this
inflationary scenario that most of the present inertia (matter) and entropy
is created. In turn, it is precisely the growth of such inertia (i.e. $%
K\rightarrow\frac{1}{3}$) that would lead the universe to a graceful exit
from inflation. The inflationary scenario would probably last as long as it
would take for the dissipation parameter to grow to $K\rightarrow\frac{1}{3}$%
. Thus, in this model, inflation and its immediate self destruction are
natural initial conditions to the current evolution of the universe.

\section{Consequences of vacuum decay}

\subsection{Evolution of the matter fields density, $\protect\rho_{m}\left(
a\right) $}

The preceding analysis for the evolution of $\rho _{v}$ has been based on
the source equation (Eq. 2.7). In order to discuss the evolution of the
matter fields density $\rho _{m}\left( a\right) $ in the presence of a
creation pressure $\pi _{c}$, the above results can be introduced either in
the energy balance equation (Eq. 2.4) or the sink equation (Eq. 2.8).
Choosing the latter and rewriting Eq. 2.4 as a function of $a\left( t\right) 
$ one finds 
\begin{equation}
a\acute{\rho}_{m}+3\gamma \rho _{m}+3\pi _{c}=0,  \tag{4.1}
\end{equation}%
where we have used Eq. 2.10 and $\theta =3H$. From Eqs. 3.10, 
\begin{equation}
3\pi _{c}=2\left( 1-2\sin ^{2}\psi \right) \rho _{v},  \tag{4.2}
\end{equation}%
where $\rho _{v}$ is given by Eq. 3.14. Introducing these results in Eq. 4.1
gives 
\begin{equation}
a\acute{\rho}_{m}+3\gamma \rho _{m}+2\left( 1-2\sin ^{2}\psi \right) \left[ 
\bar{\beta}\left( \frac{\alpha }{a}\right) ^{\left( 2+\sin 2\psi \right) }%
\right] =0,  \tag{4.3}
\end{equation}%
where, as previously established, $\frac{d\psi }{da}=\left[ a\ln \left( 
\frac{\alpha }{a}\right) \right] ^{-1}$. Eq. 4.3 governs the evolution of
the matter fields density $\rho _{m}\left( a\right) $ in the model.

In this article the main aim has been to build a framework for discussing
both the Cosmological Constant Problem and point a way to discussing the 
\textit{dynamical problem}. In the previous section we have touched on the
question of \ \textquotedblleft why $\Lambda$ is small
now\textquotedblright. It still remains to discuss the remaining problems.
As it turns out, the results of the preceding section are sufficient for
such a discussion. Consequently, we defer a discussion of the solutions to
Eq. 4.3 to concentrate on the two remaining problems.

\subsection{The coincidence problem and the dynamical problem}

Using the results of Eqs. 3.10 to 3.12 one finds that the vacuum dilution
rates relate directly to the vacuum to matter ratios $\beta $ in the
universe. In particular, at $\sin ^{2}\psi =0$,\ when the vacuum is at its
most dilution rate, Eq. 3.11 constrains the minimum density value of the
vacuum to $_{\min }\rho _{v}=\left( \frac{3K}{6K+2}\right) \left( 3\gamma
-2\right) \rho _{m}.$ On the other extreme at $\sin ^{2}\psi =1$,\ when the
vacuum is at its least dilution rate, Eq. 3.12 constrains the maximum
density value of the vacuum to $_{\max }\rho _{v}=\left( \frac{3K}{6K-2}%
\right) \left( 3\gamma -2\right) \rho _{m}.$ These results imply that as the
universe evolves, the vacuum and matter fields are coupled through the
coupling parameter $K$ (see Eq. 3.16). The vacuum then tracks the matter
fields naturally within the bounds given by Eqs. 3.11 and 3.12 as 
\begin{equation}
\left( \frac{3K}{6K+2}\right) \left( 3\gamma -2\right) \rho _{m}\leq \rho
_{v}\leq \left( \frac{3K}{6K-2}\right) \left( 3\gamma -2\right) \rho _{m}. 
\tag{4.4}
\end{equation}%
In particular, during the radiation era $\gamma =\frac{4}{3}$, the vacuum
oscillates between the values \ $\left( \frac{3K}{3K+1}\right) \rho _{m}\leq
\rho _{v}\leq \left( \frac{3K}{3K-1}\right) \rho _{m}$. On the other hand,
during the cold matter era $\gamma =1$, the vacuum oscillates between the
values $\left( \frac{3K}{6K+2}\right) \rho _{m}\leq \rho _{v}\leq \left( 
\frac{3K}{6K-2}\right) \rho _{m}$. Clearly, provided $K>\frac{1}{3}$, the
vacuum and matter field densities, $\rho _{v}$ and $\rho _{m}$ will track
each other naturally. Further, as pointed out in Section. 3 the one-way
character of the dissipation process implies that as the universe gets
older, the upper bound on $K$ grows so that $\beta _{\max }\longrightarrow 1$
in the far future as $t\longrightarrow \infty $. Thus, as long as the
universe expands, the vacuum energy and matter density fields track each
other with decaying amplitudes. This is our explanation of the so-called
current \textit{coincidence problem}. Further, as the universe gets older,
the field oscillation amplitudes become smaller and also more stretched out
since in this case $a>>a_{0}\Longrightarrow $ $\sin \psi \left( a\right)
\longrightarrow 0$. As a result the distant future universe creates
conditions under which $\left[ \sigma \left( \psi \right) =2+\sin 2\psi
\left( a\right) \right] \longrightarrow 2$. Under these conditions $\sigma $
can be replaced by its average value, $<\sigma >\,=2$.

Finally, since the dynamics of the universe is determined by the behavior of
the fields therein, the foregoing results can be used to predict the future
evolution of the universe. In this sense, the results also address the 
\textit{dynamical problem}.

\subsection{Observational tests}

The model predicts that the universe undergoes periods of variable
(quasi-periodic) acceleration. Thus the observed current cosmic acceleration
is one of these phases. In this respect the local (in time) dynamical
evolution of the universe has two possibilities (Eq. 3.10): the universe is
either moving away or towards a local $_{\max }\rho _{v}$. An increase in
acceleration as a function of redshift $z$, for example, would indicate the
universe is moving away from a local maximum of vacuum domination $\beta
_{\max }$, and vise versa. The effective equation of state, in Eq. 3.15
provides information about this variability and how the resulting motion in
the current local phase deviates from that due to a (would-be) constant
background $\Lambda $ vacuum. This behavior can be tested by future\
experiments like the SNAP project [17]. Further, a signature of such
oscillations should be imprinted on CMB. A consideration of these
observables is ongoing and will be reported in future.

\section{Conclusion}

In this paper a framework is proposed for addressing both the Cosmological
Constant Problem and the associated \textit{dynamical problem}. The
underlying premise is that the background dark energy is both dynamical and
interacting. Such a feature gives rise to a universe that behaves as a
cosmic thermodynamic engine, with the vacuum (dark) energy as the input
fuel. This energy does work by accelerating the universe. However, like any
engine, it is impossible for the universe to use all the fuel (vacuum
energy) to do work without some of it being dissipated. The rationale for
such vacuum energy dissipation and the implied increased inertia (matter
creation) is embodied in a \textit{Cosmic Equilibrium\ Conjecture} we make
with regard to the need, on the part of the universe, to always seek for
equilibrium conditions. The increase in inertia is facilitated by a creation
pressure $\pi_{c}$ that arises as a back-reaction to the cosmic
acceleration. The treatment, based on \textit{Cosmic Equilibrium\
Conjecture, }provides a basis for a first law of cosmic dynamics which
naturally gives rise to the second (entropy law) of cosmic dynamics.

\smallskip As pointed out in Section 2, \ the \textquotedblleft why is $%
\Lambda $ small now?\textquotedblright\ part of the Cosmological Constant
Problem is addressed by noting that in this model $<\sigma >=2$, so that $%
<\rho _{v}>$ evolves as $\sim a^{-2}$. Such inverse square behavior was
established in Section 3 From the inflation era to date, $a(t)$ has evolved
by about $10^{60}$. This implies that $\Lambda _{now}\approx 10^{-60<\sigma
>}\Lambda _{pl}=10^{-120}\Lambda _{pl}$, a result which is consistent with
observations.

It was discussed, in some detail, how the vacuum energy density $\rho_{v}$
couples to the matter fields $\rho_{m}$ through matter creation pressure $%
\pi_{c}$. This coupling is facilitated by a parameter $K$ which to date (for
the post-inflationary period) we have constrained to $\frac{1}{3}<K\lesssim%
\frac{4}{9}$. The process gives rise to a cosmic vacuum energy density which
oscillates with a decaying amplitude. In turn, the coupled matter fields
oscillate in sympathy. Consequently, the two coupled fields track each other
naturally, as they evolve in time. We have put bounds (Eq. 4.4) on the
relative evolution of the magnitudes of these fields up to the constrained
free parameter $K$. Because vacuum dissipation is a one-way process, in this
treatment, it follows that as the universe grows older the extremum values
of the fields $_{\max}\rho_{v}$ and $_{\min}\rho_{m}$ become increasingly
comparable so that $\beta_{\max}\longrightarrow1$. It is in this sense that
the model addresses the \textit{coincidence problem} and predicts that in
future acceleration periods the fields will become even more comparable. The
bounds put on the relative magnitudes of these fields (Eq.4.4) also imply
that the future \ evolution of these fields is predictable. Further, since
it is these same fields that drive the universe, in the first place, the
result is that the future dynamics of the universe becomes equally
predictable. In this way the model addresses the \textit{dynamical problem.}

It is pointed out that the coupling parameter $K$\ must have, at one time,
had to grow from zero to its minimum operative value $\frac{1}{3}$. This
growth corresponds to the cosmic vacuum energy changing from a purely\textit{%
\ }potential form at $K=0$\ to a partially dissipated form $K>0$. As long as 
$K<\frac{1}{3}$, there would be little or no matter created and the
universe, essentially, behaves as a near-perfect engine with $\sim100\%$\
efficiency. The vacuum energy is then, mostly, in the form of a cosmological
constant and the universe must inflate. Consequently, our treatment requires
inflation as a natural initial condition, both for creation and for the
current classical dynamics of the universe. Moreover, the growth of the
dissipation parameter $K$ in the early universe (and hence that of the
creation pressure) as a back-reaction to the inflationary acceleration,
creates a natural graceful exit out of inflation through increase of the
universe's internal energy. Thus, in this model, the\textit{\ Cosmic
Equilibrium Conjecture }implies that inflation (through dissipative
processes) oversees its own (almost immediate) destruction, to end almost
immediately. It is in this sense that the approach predicts both inflation
and a graceful exit, while at the same time justifying matter creation.
Moreover, using the same mechanism, the universe enters and eventually exits
from any subsequent accelerations, including the current one.

The model proposed here makes testable predictions that the dynamics of the
universe is predictable with a quasi-periodic character that can be verified
by future experiments. There are several issues that the model raises which
are under consideration for future report. They include, among others, a
discussion of the background global geometry as a solution of the
gravitational field equations and a consideration of effects of \ the model
on CMB.

\begin{acknowledgement}
This work was made possible by funds from the University of Michigan.
\end{acknowledgement}


\begin{thebibliography}{99}
\bibitem{[1]} A. Riess, et al, AJ, 116,1009(1998); S. Perlmutter, et al,
ApJ. 517 565(1998).

\bibitem{[2]} G. Efstathiou, MNRAS, 330 29(2002).

\bibitem{[3]} R. Stompor, Apj. 561L, 7(2001); L. M. Rebull et al,
astro-ph/0203384(2002).

\bibitem{[4]} A. Guth, Phys. Rev. D 23, 347(1981).

\bibitem{[5]} S. Weinberg, Rev. Mod. Phys. 61, 1 (1989); S. M. Carroll,
astro-ph/0004075.

\bibitem{[6]} J. M. Overduin and F. I. Cooperstock, Phys.Rev. D 58
043506(1998); M. R. Mbonye, Int. J. Mod. Phys. A18 (2003) 811;
astro-ph/0208244.

\bibitem{[7]} J. A. S. Lima, Phys. Rev. D, 54, 2571(1996).

\bibitem{[8]} M. S. Turner and M. White, Phys. Rev. D. 56 R 4439(1997).

\bibitem{[9]} B. Rastra and P. J. E. Peebles, Phys. Rev. D \textbf{37},
3406(1988).

\bibitem{[10]} S. Coleman, Nucl. Phys. B, \textbf{307}, 867(1988).

\bibitem{[11]} R. H. Brandenberger, hep-th/0210165.

\bibitem{[12]} R. R. Caldwell, R. Dave and P. J. Steinhardt, Phys. Rev.
Lett. \textbf{80}, 1582(1998).

\bibitem{[13]} J. Carriaga, M. Livio and A. Valenkin, Phys. Rev. D \textbf{61%
}, 023503(2000); S. B. Bludman, Nucl. Phys. A 663, 865(2000).

\bibitem{[14]} (a) T. Padmanabhan, Class. Quantum. Grav., \textbf{19, }%
5387(2002); (b) T. Padmanabhan, hep-th/0212290.

\bibitem{[15]} J. D. Jackson, \textit{Classical Electrodynamics, John Wiley
and Sons, New York (1975).}

\bibitem{[16]} J. W. Rohlf, \textit{Moderen Physics from a to Z0, Wiley
(1994). }

\bibitem{[17]} A. Kim, (for SNAP Collaboration), astro-ph/0210077; G. Tarle,
et al (for SNAP collaboration), astro-ph/0210041.

\bibitem{[18]} P. J. E Peebles, \textit{Principles of Cosmology, Princeton
Univ. Press, Princeton NJ} \textit{(1993)}; David Tytler et al,
astro-ph/0001318(2000).

\bibitem{[19]} R. Silva, \ J. A. S. Lima and M. O. Calv\~{a}o, gr-qc/0201048
(2002)

\bibitem{[20]} M. O. Calvao, J. A. S. Lima and I. Waga, Phys. Lett. A 162,
223(1992).

\bibitem{[21]} J. Gaheniau, E. Gunzig and I. Stengers, Found.of. Phys. 17,
585(1987); S. R. de Groot and P. Mazur, \textit{Non-Equilibrium
Thermodynamics, Dover, NY, USA (1984).}
\end{thebibliography}
\end{document}